\documentclass[twocolumn,showpacs,preprintnumbers,amsmath,amssymb]{revtex4}
\usepackage{amsmath,amssymb,color,epsfig,latexsym,natbib,graphicx,dcolumn,ulem}
\newcommand{\bm}{\mathbf}
\bibpunct{[}{]}{,}{n}{,}{,}             
\def\DEL#1{{\textcolor{green}{}}}         

 \newcommand{\bB}{\bf{B}}

\newcommand{\rem}[1]{}

\newcommand\vecp[1]{\vec{#1}}                   

\newcommand{\be}{\begin{equation}}
\newcommand{\ee}{\end{equation}}

\def\bB0{\vecp{B}_0}

\begin{document}
\title{\bf On the lack of universality in decaying magnetohydrodynamic turbulence} 

\author{E. Lee$^{1,2}$, M.E. Brachet$^{1,3}$, A. Pouquet$^1$, P.D. Mininni$^{1,4}$, and D. Rosenberg$^1$}
\affiliation{(1) Geophysical Turbulence Program, National Center for Atmospheric Research, P.O. Box 3000, Boulder, Colorado 80307-3000, USA }
\affiliation{(2) Department of Applied Physics and Applied Mathematics,
Columbia University, 500 W. 120th Street, New York NY 10027, USA }
\affiliation{(3)  \'Ecole Normale Sup\'erieure, 24 rue Lhomond, 75005 Paris, France}
\affiliation{(4) Departamento de F\'\i sica, Facultad de Ciencias Exactas y
         Naturales, Universidad de Buenos Aires, Ciudad Universitaria, 1428
         Buenos Aires, Argentina}
         
\begin{abstract}
Using computations of three-dimensional magnetohydrodynamic (MHD) turbulence with a Taylor-Green flow, whose inherent time-independent symmetries are implemented numerically, 
and in the absence of either a forcing function or an imposed uniform magnetic field, we show that three different inertial ranges for the energy spectrum may emerge for three different initial magnetic fields, the selecting parameter being the ratio of the Alfv\'en to the eddy turnover time.  Equivalent computational grids range from $128^3$ to $2048^3$ points with a unit magnetic Prandtl number and a Taylor Reynolds number of up to 1500 at the peak of dissipation.  We also show convergence of our results with Reynolds number. 
 Our study is consistent with previous findings of a variety of energy spectra in MHD turbulence by studies performed in the presence of both a forcing term with a given correlation time and a strong, uniform magnetic field.
In contrast to the previous studies, however, the ratio of characteristic time scales here can only be ascribed to the intrinsic nonlinear dynamics of the flows under study.
\end{abstract}
\maketitle

\section{Introduction} \label{s:intro}

Turbulence forms the backbone of many natural phenomena in the atmosphere and ocean, as well as in astrophysical flows.  In the latter, it is often accompanied by the coupling of vortices and current structures.  For incompressible neutral fluids, under the assumption of a high Reynolds number (and therefore a long dissipation time), as is the case for many geophysical flows, the relevant time scale for the problem is the nonlinear eddy turnover time.  For such flows, the phenomenology developed by Kolmogorov in 1941 (hereafter referred to as ``K41'') \cite{K41}, predicting a kinetic energy spectrum $E^V_{K41}(k)\sim k^{-5/3}$, represents a good first approach even if corrections to this phenomenology for higher-order statistics are known to exist, due to the breakdown of the self-similarity represented by a simple power-law energy spectrum.  When electromagnetic forces are introduced, other time scales can arise, such as the Alfv\'en time, associated with the propagation of transverse waves along magnetic field lines.  K41 phenomenology may still apply, but one must also consider the role of Alfv\'en waves in producing a different power law for the total energy spectrum, as illustrated independently by Iroshnikov and Kraichnan (hereafter, ``IK'') \cite{IK1}:  $E^T_{IK}(k)\sim k^{-3/2}$.

Isotropy is assumed by both K41 and IK, but it is not necessarily achieved.  In neutral flows, if the anisotropy of the small scales, in the form of elongated vortex filaments, occurs locally in space, isotropy may be recovered overall because the filaments are randomly oriented, and the vorticity spectrum $k^2E^V(k)$, which peaks in the small scales, contributes little to the large-scale spectrum.  In contrast, the anisotropy of MHD originates from a large-scale magnetic field, which can be dominant energetically and relevant at all scales.  Studies of anisotropic MHD date back to the mid-fifties for liquid metals at low magnetic Reynolds number \cite{lehnert} (see also \cite{frisch_aniso}), and a bit later for fully developed MHD turbulence \cite{strauss, montgomery}.  More recently, a wealth of new studies on MHD turbulence has been made possible 
\cite{carbone,galtier_weak3, boldyrev}  
in part by the revival of weak turbulence theory (e.g.~\cite{galtier00} for MHD), the availability of more detailed observations \cite{saur,podesta}, and improved resolution in numerical simulations \cite{mueller, 1536b, mason_08,perez,gafd}.

From the theoretical point of view, the presence of a strong background magnetic field $\bm B_0$ allows for the existence of a small parameter characterizing the ratio of velocity and magnetic field fluctuations to $|\bm B_0|$, enabling an analytical solution via the weak turbulence (hereafter, ``WT'') approach.  In contrast to the K41 and IK spectra, the WT energy spectrum is anisotropic:  $E_{WT}(k_{\perp}, k_{\parallel})\sim k_{\perp}^{-2}$, where perpendicular and parallel are relative to the direction of $\bm B_0$; there is, in fact, no prescribed transfer in $k_{\parallel}$ at lowest order.  It is interesting to note that the IK phenomenology is compatible with weak turbulence in the isotropic limit $k_{\perp} \approx k_{\parallel}$, giving it a stronger theoretical footing.  Other phenomenological approaches hypothesize that even with a strong background field $\bm B_0$ making the flow highly anisotropic, an ``anisotropic Kolmogorov'' scaling of the energy spectrum 
is possible by way of a dynamical effect that makes the two characteristic times of the problem (the Alfv\'en time and the eddy turnover time) equal at all scales \cite{goldreich_critical}.  In fact, when this hypothesis is relaxed to a constant ratio across the inertial range (not necessarily equal to unity), the dynamics can then be shown to be compatible with a variety of inertial range scalings, including K41, IK and WT \cite{galtier_weak3}. Weak MHD turbulence has been observed in the magnetosphere of Jupiter \cite{saur}, where the strong Jovian field creates a favorable environment for wave interactions to dominate the dynamics.  In the solar wind, data for a long time have indicated that the spectrum appears Kolmogorovian \cite{matthaeus_SW}, although recent observations indicate a more complex dynamics (discussed further in the conclusion, \S \ref{s:con}).

Numerical simulations to date are unable to give a definitive answer to the question of spectral index in plasma turbulence, at least in three dimensions.  The difference between the K41 and IK scalings is subtle enough that any type of contamination, in particular by intermittency as well as dissipative small-scale effects, will blur the results.  Intermittency, the sporadic occurrence of intense small-scale structures, tends to steepen the direct cascade energy spectrum.  In fact, it has been shown both in two and three spatial dimensions that intermittency in MHD generally leads to stronger corrections to high-order structure functions than in neutral fluids \cite{intermi_MHD2d, intermi,1536b}, and that the magnetic field is more intermittent than the velocity \cite{intermi_MHD2d,1536_mhd_new}.  

Recent studies indicate that, in the presence of an external force with a given autocorrelation time and/or a strong, uniform magnetic field, the energy spectrum can exhibit different power laws (e.g., \cite{matthaeus_RMHD, rapazzo_07,mason_08}).  Such varied spectral indices can be ascribed to the variation of time scales \cite{mueller} or to the presence of complex structures, such as ribbons (see \cite{mason_08} and references therein, \cite{barbara}). 
However, the possibly simpler problem of incompressible MHD decay in three dimensions with $\bm B_0 = 0$ has not been examined in this light \cite{note1}.  Therefore, it is the purpose of this paper to do so by way of high-resolution numerical simulation and to show that indeed several classes of dynamics are possible in decaying MHD turbulence.  In the next section, we give equations and initial conditions; \S \ref{s:reg} is dedicated to the temporal behavior of the flows, \S \ref{s:sca} to the spectra observed in this paper and convergence of the results with Reynolds number, and \S \ref{s:con} to
a discussion and brief concluding remarks.

\section{Three classes of Magnetic Taylor-Green flow}\label{s:equ}

The MHD equations for an incompressible flow with a velocity $\bm v$ and magnetic induction $\bm b$ (in units of Alfv\'en velocity) read:
\begin{equation}
\frac{\partial \bm v}{\partial t} + \bm v \cdot \nabla \bm v = -\nabla p + \bm j \times \bm b + \nu \Delta \bm v \ ,
\end{equation}
\begin{equation}
\frac{\partial \bm b}{\partial t} = \nabla \times (\bm v \times \bm b) + \eta \Delta \bm b \ ,
\end{equation}
\[
\nabla \cdot \bm v = 0 = \nabla \cdot \bm b \ ,
\]
with $p$ the fluid pressure and $\bm j = \nabla \times \bm b$ the current density.
In the absence of viscosity $\nu$ and resistivity $\eta$, the total energy
$E^T=\langle{\textbf{v}}^2+\bm b^2\rangle /2 =E^V+E^M$, cross helicity
$H^C=\langle\bm v \cdot \bm b\rangle /2$, and magnetic helicity 
$H^M= \langle\textbf{a} \cdot \textbf{b}\rangle /2$ (where $\bm b \equiv \nabla \times \bm a$ defines the magnetic potential, $\bm a$) are conserved. 
The Reynolds number here is defined as $Re=v_{rms}L^T/\nu$, and the magnetic Reynolds number as $Rm=v_{rms}L^M/\eta$, with the integral scale and kinetic and magnetic integral scales defined, respectively, as:  
$$
L^{V,M,T} = 2\pi \frac{\int{E^{V,M,T}(k) k^{-1} dk}}{\int{E^{V,M,T}(k) dk}} \ .
$$

Similarly, one can define the Taylor Reynolds number
$R_{\lambda}=v_{rms}\lambda^T/\nu$, where the Taylor scale $\lambda^T$ is defined as
$$
\lambda^T = 2\pi \left(\frac{\int{E^T(k) dk}}{\int{E^T(k) k^2 dk}}\right)^{1/2} \ .
$$
Kinetic and magnetic Taylor scales can also be defined:
$$\lambda^{V,M} = 2\pi \left(\frac{\int{E^{V,M}(k) dk}}{\int{E^{V,M}(k) k^2 dk}}\right)^{1/2} \ .$$

Ideal MHD ($\nu=0$, $\eta=0$) has been studied numerically both in two dimensions \cite{ideal_2d} and in three \cite{kerr,first}, including with adaptive mesh refinement \cite{grauer}.  Such simulations are important for understanding the initial nonlinear exchanges among modes until the smallest resolved scales are reached, at which point dissipation must be introduced to continue the computation and reach a fully developed turbulent flow with current and vorticity sheets.

The velocity field we choose for our initial conditions is the Taylor-Green (hereafter, ``TG'') vortex \cite{TG}  corresponding to a von K\'arm\'an flow between two counter-rotating disks.
The simplest TG velocity field can be written as \cite{brachet83} (see also \cite{brachet90}):
\begin{equation}
\bm v_{TG}(x, y, z) = v_0 \left[ \sin x \cos y \cos z \ \bm{\hat{e}_x} - \cos x \sin y \cos z \ \bm{\hat{e}_y} \right] .  
\nonumber  \end{equation}
The velocity component in the third direction is zero initially but grows with time.  We also define
 $\boldsymbol\omega_{TG}=\nabla \times \bm v_{TG}$ (and $\boldsymbol\omega = \nabla \times {\bf v}$, the vorticity as usual). This TG vortex has been used not only in numerical studies \cite{nore}, but also extensively in laboratory studies of fluid turbulence and as a driver for the generation of magnetic fields within liquid metals flows \cite{bourgoin02}.

A generalization of the TG vortex symmetries to MHD was presented in \cite{first}, where the ideal case was studied with the following initial magnetic field configuration:
\begin{eqnarray}
b_x^{I} &=& b_0^I \cos(x)\sin(y)\sin(z) \\
b_y^{I} &=& b_0^I \sin(x)\cos(y)\sin(z) \\
b_z^{I} &=& -2 b_0^I \sin(x)\sin(y)\cos(z) .
\label{eqn:btg_I}\end{eqnarray}

It can be shown that the magnetic field $\bm b^I$  is everywhere perpendicular to the faces, or ``walls,'' of a box defined as $[0,\pi]^3$.  The current $\bm j^I = \nabla \times \bm b^I$ is then found to be parallel to the walls, which thus can be considered as electrical insulators; for this reason, we refer to this type of TG flow as ``insulating.''  Also of interest is that the cross-correlation $H_C$ between the velocity and magnetic fields is identically zero globally, and $\bm b^{I} = -[b_0/v_0] \boldsymbol\omega_{TG}$.

Another magnetic field was proposed in \cite{first}, namely:
\begin{eqnarray}
b_x^{C} &=& = b_0^C \sin(2x)\cos(2y)\cos(2z) \\
b_y^{C} &=& = b_0^C \cos(2x)\sin(2y)\cos(2z) \\
b_z^{C} &=& = -2 b_0^C \cos(2x)\cos(2y)\sin(2z) \ ,        \label{eqn:btg_C}\end{eqnarray} 
where the current in the $[0,\pi]^3$ box here is perpendicular to the walls, which are therefore ``conducting.''  In this configuration, $H^C$ is non-zero but weak (less than $4\%$ at its maximum over time, in a dimensionless measure relative to the total energy).

Finally, we also introduce an alternative to the insulating magnetic induction above, which we call $\bm b^A$ (for ``alternative'' insulating flow), defined as:
\begin{eqnarray}
b_x^{A} &=& = b_0^A \cos(2x)\sin(2y)\sin(2z) \\
b_y^{A} &=& = -b_0^A \sin(2x)\cos(2y)\sin(2z) \\
b_z^{A} &=& = 0 \ ,
\label{eqn:btg_A}\end{eqnarray}
configuration for which again $H_C=0$. Note that the magnetic helicity is zero in all three configurations.

All flows are initialized at the largest scales in order to obtain the highest possible Reynolds number, and all have unit magnetic Prandtl number (i.e., $\nu=\eta$).  The values of the parameters for all runs described in this paper are given in Table \ref{table1}.  For the three classes of flow proposed here (named hereafter ``I,'' ``A,'' and ``C,'' referring to the three induction configurations), the same TG velocity is applied at $t=0$, and the fields are normalized such that $E^V(t=0) = E^M(t=0) = 0.125$: at the start of each simulation, the kinetic and magnetic energies are in equipartition.  The resolutions of the runs range from $64^3$ to $2048^3$ points (in terms of equivalent grids for computations that would not exploit the symmetries; see \cite{first}), allowing the range of Reynolds number to vary over a factor close to 40.

Taylor-Green configurations possess several inherent symmetries within a cube of length $2\pi$ (periodicity).
 Mirror symmetries about the planes $x=0$, $x=\pi$, $y=0$, $y=\pi$, $z=0$, and $z=\pi$; 
and rotational symmetries of angle $N\pi$ about the axes $(x,y,z)=(\frac{\pi}{2},y,\frac{\pi}{2})$ and $(x,\frac{\pi}{2},\frac{\pi}{2})$ and of angle $N\pi/2$ about the axis $(\frac{\pi}{2},\frac{\pi}{2},z)$, for $N \in \mathbb{Z}$.  All flows are defined in the $[0,2\pi]^3$ box and satisfy periodic boundary conditions of the domain.  Within the domain, the planes of mirror symmetry mentioned above form the insulating and conducting walls of the smaller $[0,\pi]^3$ boxes.

\begin{figure}[htb]
\includegraphics[width=7cm, height=50mm]{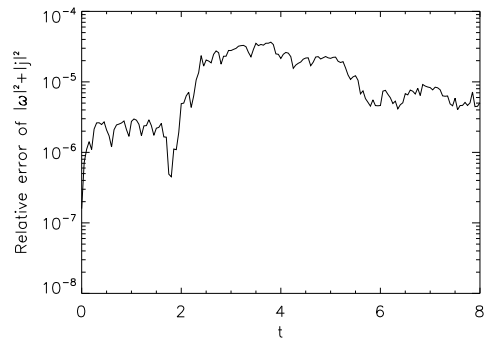}
\caption{
Relative error in the total enstrophy as a function of time for the I3 flow ($Re = 940$) between a full direct numerical simulation on a grid of $512^3$ points with no imposed symmetries and one making use of code-implemented symmetries using the same grid.
}
\label{f_error} \end{figure}

As the symmetries of the TG flows are also symmetries of the MHD equations, they are preserved by time evolution of the solutions.  Numerical implementation of the symmetries allows for substantial savings in both computing time and memory usage at a given Reynolds number, with no approximation or closure scheme needed. The numerical method is pseudo-spectral, with minimum wavenumber $k_{min}=1$ and maximum wavenumber $k_{max}=N/3$, where $N$ the number of grid points in each direction, using a 2/3-deliasing rule; the temporal integration is performed using a second-order Runge-Kutta scheme.  The code follows the parallelization, using MPI, as described in \cite{pablo_code1}
(see \cite{brachet83} for a detailed implementation for Euler and Navier-Stokes and \cite{first} for MHD). 
It was checked for runs at a resolution of $512^3$ grid points that the differences between results obtained with the code implementing all the symmetries and those with a general code (in which the symmetries are not imposed explicitly, but the initial conditions are the same) were small and did not grow in time: in Fig.~\ref{f_error} is shown the relative difference for the domain-integrated total enstrophy, defined as  
$\langle \omega^2 + j^2 \rangle/2$, which remains of the order of $10^{-5}$ throughout the computation (see also \cite{first}). Further results concerning such a comparison, as well as an analysis of the overall dynamics of the I6 flow are given in \cite{gafd}.


\begin{figure}[htb]
\includegraphics[width=7cm, height=50mm]{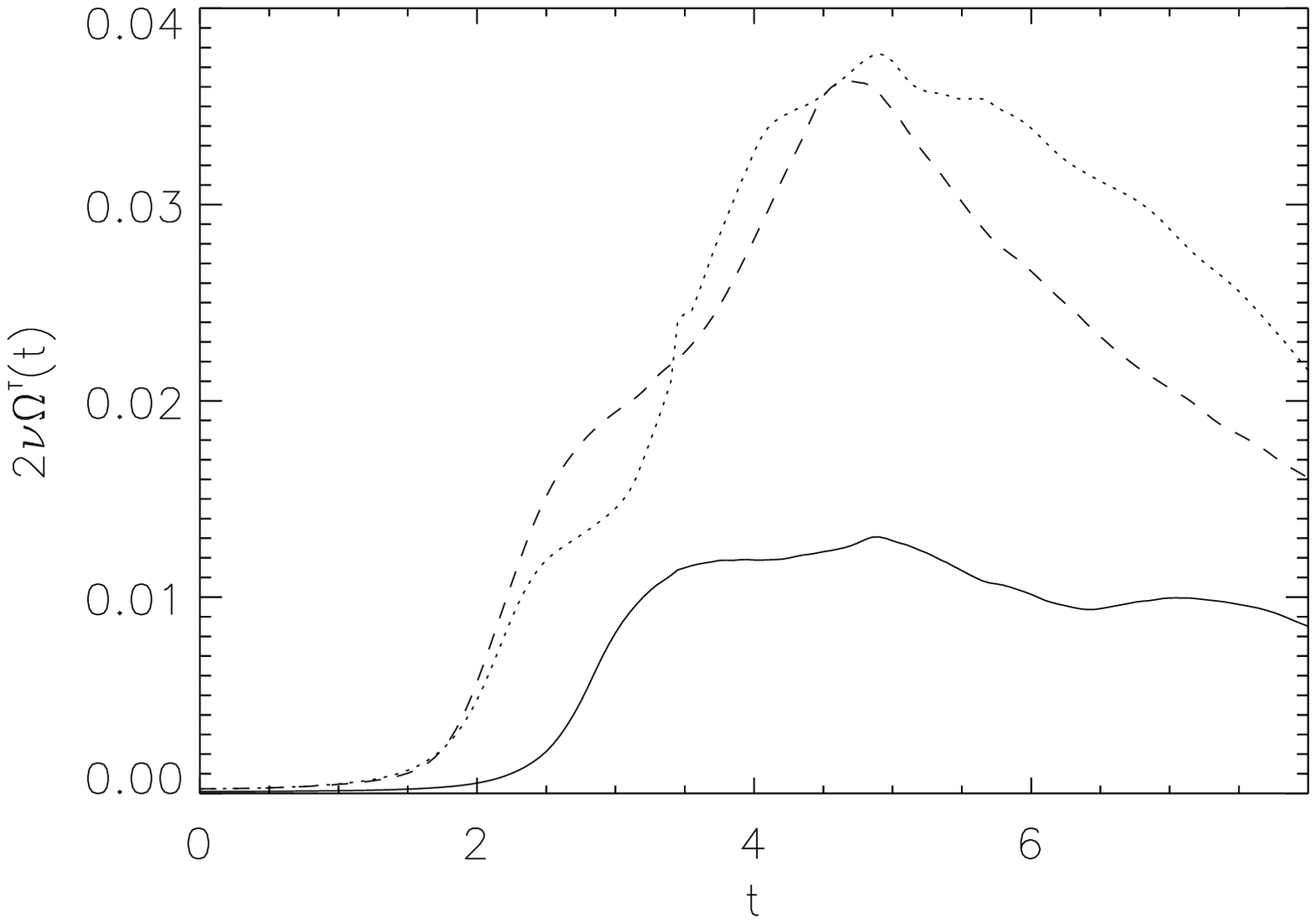}\\
\includegraphics[width=7cm, height=50mm]{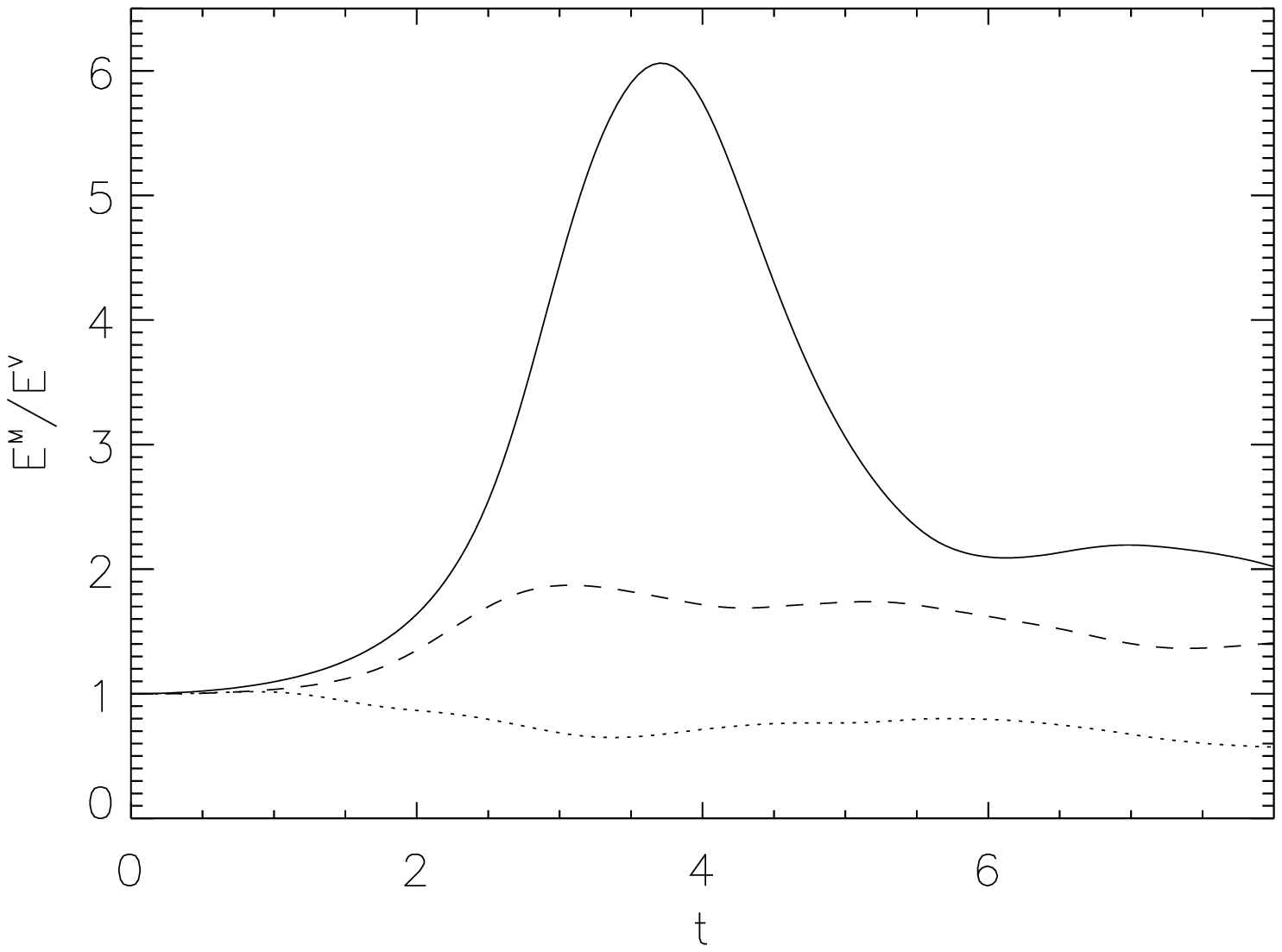}
\caption{ 
(\textit{Top}) Temporal evolution of the total dissipation for the highest Reynolds numbers for each type of flow, represented by runs I6 (solid), A6 (dashed), and C6 (dotted), as described in Table \ref{table1}.  (\textit{Bottom}) Ratio of total magnetic to kinetic energy $E^M/E^V$ for these same runs.  Note that I6 has noticeably less dissipation and more magnetic energy.}
\label{f_temp_enr_diss} \end{figure}


\begin{table}[t] \caption{\label{table1}
Parameters used in the simulations. $N$ is the equivalent linear grid resolution; $v_{rms}$ and $b_{rms}$ are the rms velocity and magnetic field at peak of dissipation; $\nu=\eta$ is the kinematic viscosity; and $Re$ is the Reynolds number based on the integral scale of the flow.
Note the growth of $b_{rms}$ with $Re$ in all but one example, as well as the diminishing ratio $b_{rms}/v_{rms}$ going from I to A to C flows.
$^{\ast}$ I3 has also been run on a full grid, for comparison purposes (see Fig. \ref{f_error}).
}
\begin{ruledtabular} \begin{tabular}{cccccc}
Run & N& $v_{rms}$ & $\nu$ & $b_{rms}$ & $Re$  \\
\hline
I1 & 128 &  0.27  &$2\times 10^{-3}$& $0.58$ &  $250$    \\
I2 & 256 &  0.27  &$1\times 10^{-3}$& $0.60$ &  $490$    \\
\ I3$^{\ast}$ & 512 &  0.27  &$5\times 10^{-4}$& $0.62$ &  $940$    \\ 
I4 & 512 &  0.27  &$2.5\times 10^{-4}$& $0.63$ &  $1800$     \\ 
I5 & 1024 &  0.27  &$1.25\times 10^{-4}$& $0.63$ &  $3400$    \\ 
I6 & 2048 &  0.32  &$6.25\times 10^{-5}$& $0.59$ &  $9700$   \\ 
  \hline
A1 & 64 &  0.39  &$2\times 10^{-3}$& $0.43$ &  $260$    \\
A2 & 128 &  0.38  &$1\times 10^{-3}$& $0.46$ &  $460$    \\
A3 & 256 &  0.37  &$5\times 10^{-4}$& $0.46$ &  $780$    \\ 
A4 & 512 &  0.37  &$2.5\times 10^{-4}$& $0.47$ &  $1500$     \\ 
A5 & 1024 &  0.37  &$1.25\times 10^{-4}$& $0.48$ &  $2900$    \\ 
A6 & 2048 &  0.37  &$6.25\times 10^{-5}$& $0.49$ &  $5600$   \\ 
\hline
C1 & 64 &  0.49  &$2\times 10^{-3}$& $0.31$ &  $460$    \\
C2 & 128 &  0.47  &$1\times 10^{-3}$& $0.32$ &  $810$    \\
C3& 256 &  0.46  &$5\times 10^{-4}$& $0.34$ &  $1500$    \\ 
C4 & 512 &  0.46  &$2.5\times 10^{-4}$& $0.37$ &  $2700$     \\ 
C5 & 1024 &  0.46  &$1.25\times 10^{-4}$& $0.39$ &  $4900$    \\ 
C6 & 2048 &  0.45  &$6.25\times 10^{-5}$& $0.40$ &  $9100$   \\ 
\end{tabular} \end{ruledtabular} \end{table}


\begin{figure}[t]
\includegraphics[height=60mm]{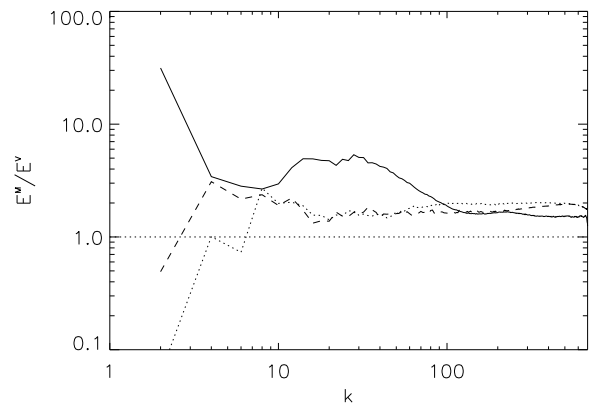}
\caption{
Ratio of magnetic to kinetic energy spectra averaged over $\Delta t = 0.5$ (1.5 to 2 turnover times) about the maximum of dissipation for run I6 (solid), A6 (dashed), and C6 (dotted).  (Labels are the same as in Fig.~\ref{f_temp_enr_diss}; see Table \ref{table1} for run details.)
Note the dominance of the magnetic energy at large scale for run I6 (solid), and the tendency towards equipartition at small scales for all runs, with a slight excess of magnetic energy.
} 
\label{f_spec_ratio_enr} \end{figure}

\begin{figure}[t]
\includegraphics[width=7cm, height=50mm]{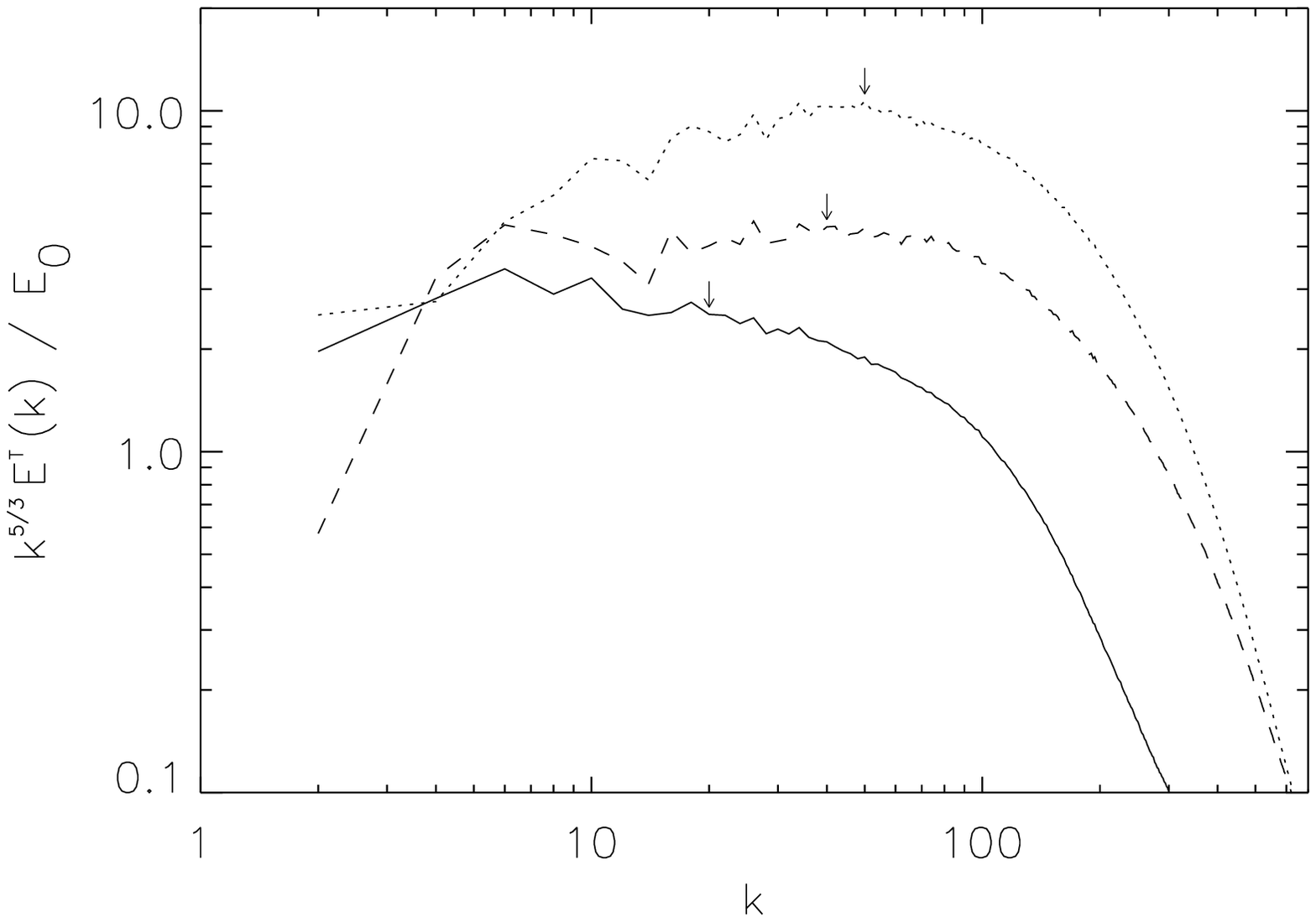}\\
\includegraphics[width=7cm, height=50mm]{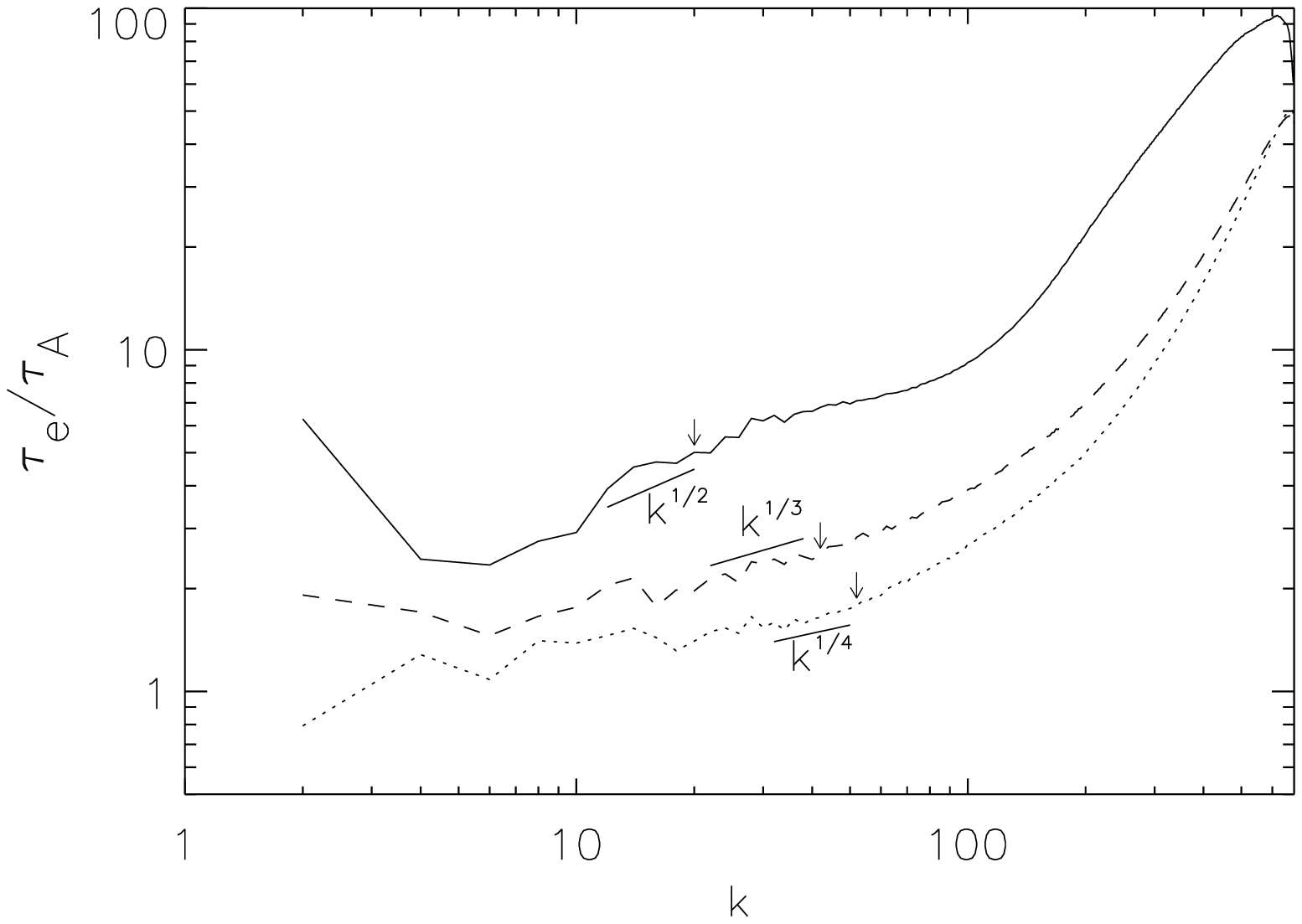}
\caption{ 
Total energy spectra (\textit{top}) compensated by $k^{5/3}$ and averaged over $\Delta t = 0.5$ (1.5 to 2 turnover times) about the maximum of dissipation, and ratio of nonlinear to Alfv\'en time scales as a function of wavenumber (\textit{bottom}) for the same runs (labels as in Fig.~\ref{f_temp_enr_diss}: solid line for I, dash for A and dots for C).  Slopes are given only as a reference.  The three arrows indicate the magnetic Taylor scale. Note that the three spectra follow noticeably different spectral laws and possibly different scale-dependence for their time scales as well (see text).}
\label{f_spectra_enr_times} \end{figure}


\section{The emergence of different regimes} \label{s:reg}

In a three-dimensional non-rotating turbulent flow, the nonlinear terms---here, the momentum advection, Lorentz force, and induction terms---produce coupling between modes, and both the kinetic and magnetic energy are transferred to small scales as a result.  As progressively smaller scales become excited, the volume-integrated vorticity and current density grow until dissipation sets in, as can be seen in Fig.~\ref{f_temp_enr_diss} (\textit{top}) for the highest-resolution runs I6 (solid line), A6 (dashed line) and C6 (dotted line), following the nomenclature of Table \ref{table1}.
The total level of dissipation is comparable for runs A6 and C6 but substantially lower for run I6. 
The ratios of magnetic to kinetic dissipation are smaller than their equivalent energy counterparts, and they reach their maxima earlier (not shown); this indicates a stronger tendency towards equipartition in the small scales, as can be expected since the Alfv\'en time varies as $1/k$ (the data for run I6 is discussed in \cite{gafd}).

The energy exchanges between the velocity and magnetic fields are complex, yielding different ratios of magnetic to kinetic energy that change both in time and from flow to flow, as we see in Fig.~\ref{f_temp_enr_diss} (bottom).  These differences can also be understood in terms of the diversity of nonlinear terms in the MHD equations, their relative importance depending on the initial conditions.  For example, runs I are dominated by magnetic energy after a short period of time, although the initial fields are in equipartition.  For this flow, the magnetic field and the vorticity are initially parallel at every point in space.  As a result, the nonlinear terms in the MHD equations initially lead to a more rapid production of magnetic energy than in the other flows.  At later times, the action of the Lorentz force differentiates the evolution of the magnetic field from that of the vorticity.

We find thus that the intrinsic nonlinear dynamics for the three sets of initial conditions lead to magnetically dominated (I) flows, quasi-equipartitioned (A) flows, or flows with sub-dominant magnetic energy (C), at least in the large scales.  Fig.~\ref{f_spec_ratio_enr} displays the variation with wavenumber of the ratio of the magnetic to kinetic energy spectra for the three high-resolution runs, one for each class.  A surplus of magnetic energy at large scale for the I flow is evident, as is a tendency in all flows towards quasi-equipartition at small scales.  A slight excess of magnetic excitation at small scales can be observed, a feature perhaps linked to the absence of a magnetically induced ``eddy viscosity'' for the magnetic energy akin to one for the kinetic energy, as shown in studies of MHD turbulence using transport coefficients derived from the full integro-differential equations arising from a two-point closure \cite{PFL}.

When examining now the resulting energy spectra, averaged over an interval of time $\Delta t=0.5$ about the maximum of dissipation of each flow, one can easily distinguish the three flows, with measurably different power laws.  Fig.~\ref{f_spectra_enr_times} (\textit{top}) gives the total energy spectra for the same three runs, compensated by $k^{5/3}$, calculated from data averaged over eleven temporal outputs at $t \in [3.75,\ 4.25]$ for I6, $t \in [4.5,\ 5.0]$ for A6 and $t \in [4.75,\ 5.25]$ for C6.  The A6 flow is near the K41 scaling; the C6 flow has a shallower spectrum close to the IK dynamics, with a $k^{-3/2}$ index; and the I6 flow has a steeper spectrum close to the WT expectation, with a $k^{-2}$ power law (see also Section \ref{s:sca} ).  The denotations of the spectra as K41, IK, and WT are used here for simplicity, and as will be discussed later, more simulations are needed to decide whether these are the dynamical attractors of the equations or if other solutions exist.  Regardless, the three sets of initial conditions clearly lead to different spectral behavior, which can be linked to the several times scales involved in the system as shown next.
We recall that in \cite{1536b}, the IK spectrum was followed for larger wavenumbers by a steeper spectrum, $k_{\perp}^{-2}$ corresponding to WT.  The data in the present study are not quite sufficient to confirm this finding, but it is still highly possible that the IK spectrum in C6 (dotted line in Fig.~\ref{f_spectra_enr_times}) is followed by a steeper behavior, with a transition occurring, as in \cite{1536b}, at the magnetic Taylor scale, indicated by an arrow, whereas no such transition is visible for the other two classes, A and I.

In Fig.~\ref{f_spectra_enr_times} (\textit{bottom}) is shown the ratio of the nonlinear time to the Alfv\'en time, $R_{\tau}(k)=\tau_e(k)/\tau_A(k)$, where these times are defined as:
$$
\tau_e(k)= \frac{1}{\sqrt{2k^3E^V(k)}} \hskip0.1truein , \hskip0.1truein  \tau_A(k)= \frac{1}{kB_0} \ ,$$ \\
with $B_0$ the field in the largest--scale mode; hence:
$$
R_\tau(k) = \frac{B_0}{\sqrt{2kE^V(k)}} \ .
$$
 
Contrasting the plots in Fig.~\ref{f_spectra_enr_times}, we may conjecture that it is the competition between the two characteristic phenomena in MHD turbulence (nonlinear steepening as measured by $\tau_e$, and wave interactions as measured by $\tau_A$) that produces different equilibria among scales and therefore different energy spectra. The fact that these ratios here are always greater than unity is not significant in itself, as phenomenological determination of characteristic times leaves them within some constant factor of order unity, but what may be significant is their variation from flow to flow, as well as their variation with scale.

As discussed earlier, it has been hypothesized that MHD turbulence dynamics may be understood in the context of an equilibrium between turbulent eddies and Alfv\'en wave interactions \cite{IK1,strauss,goldreich_critical,galtier_weak3}.  Indeed, the nonlinear MHD equations accept the solutions $\bm v = \pm \bm b$.  MHD turbulence can also be viewed as the competition between nonlinear steepening and (semi)-dispersive effects, somewhat akin to soliton dynamics \cite{matthaeus_RMHD}.  The nonlinear competition between eddies and waves, in this light, perhaps could be measured by $R_{\tau}$.  By setting $R_{\tau} = 1$ at all scales, a K41-type spectrum could occur; relaxing the condition and leaving $R_{\tau}$ equal to a constant, independent of scale, a model can be devised \cite{goldreich_critical} whereby such a condition can be made compatible with the IK phenomenology and WT theory as well \cite{galtier_weak3}.

It could be argued that the hypothesis of constancy of $R_{\tau}$ with scale is verified by the data displayed in Fig.~\ref{f_spectra_enr_times}.  However, if we estimate as usual the nonlinear time as $1/\sqrt{2k^3E^V(k)}$, then $R_{\tau}$ must vary as $k^{1/3}$,  $k^{1/4}$, and $k^{1/2}$, respectively, for K41, IK, and WT dynamics.  The results shown in Fig.~\ref{f_spectra_enr_times} are also compatible with this interpretation, favoring a slight steepening of $R_{\tau}(k)$ as we go from I to A to C  initial conditions. This observation may be affected, though, by a contamination of the inertial range by dissipation associated with still somewhat moderate Reynolds numbers.  Obviously, higher Reynolds number computations will help to clarify this point.

\begin{figure}[htb]
\includegraphics[width=7cm, height=50mm]{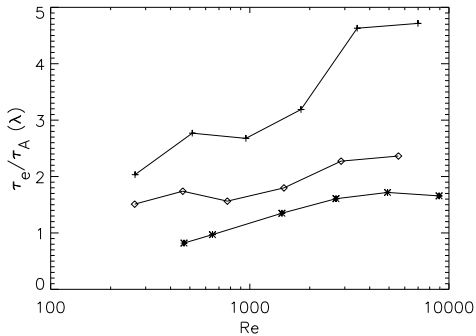}
\caption{
Variation with  Reynolds number $Re$ of the ratio of eddy turn-over to Alfv\'en time scales computed at the Taylor scale $\lambda^M$ for each flow.  I1-I6 ($+$), A1-A6 ($\diamond$), and C1-C6 ($\ast$) are plotted in order of resolution and Reynolds number, as listed in Table \ref{table1}.  All quantities are computed in an interval of $\Delta t=0.5$ about the peak of dissipation for each run.
Note the rather different values of these ratios and onset of saturation for the highest $Re$ indicative of the beginning of a convergence to a high-$Re$ regime.}
\label{f_ratio_Re} \end{figure}

\begin{table}[htb] \caption{\label{table2}
Integral scales $L^{T,V,M}$, Taylor scales $\lambda^{V,M}$, and Taylor Reynolds number $R_{\lambda}$ based on $\lambda^T$; all values are taken at the instantaneous peak of dissipation.
Note the substantially larger integral and Taylor scales in run I6, as well as the kinetic integral scale $L^V$ in C6.  These scales directly affect the Reynolds numbers of the flows and the relative time scale responsible for the dynamics.}
\begin{ruledtabular} \begin{tabular}{ccccccc}
Run & $L^T$& $L^V$ & $L^M$ & $\lambda^T$ & $\lambda^M$ & $R_{\lambda}$  \\
\hline
I6 & 1.88 & 1.75 &  1.91  & 0.29 & 0.31 &  1500   \\ 
A6 & 0.94  & 1.00 & 0.90  & 0.16 & 0.16 &  960   \\ 
C6 & 1.25 & 1.74 &  0.61  & 0.15 & 0.13 &  1100   \\ 
\end{tabular} \end{ruledtabular} \end{table}



\section{Scaling with Reynolds number} \label{s:sca}


It is notoriously difficult to measure spectral indices of power laws found in numerical simulations, in particular because of the small extent of the inertial range, sandwiched in wavenumber space between the energy-containing and dissipative scales.  A further difficulty arises when more than one inertial regime exists, as found, for example, in Hall MHD \cite{galtier_whistler} and in weak turbulence \cite{1536b}. 
 The question then arises as to whether or not the spectra plotted in Fig.~\ref{f_spectra_enr_times} and other data presented here would be consistent with the dynamics of equivalent flows at substantially higher Reynolds numbers.  To address such issues we now turn to a convergence study of the data.

In Fig.~\ref{f_spectra_enr_times}, $R_{\tau}$ is plotted as a function of scale, but to see its dependence on Reynolds number, we can compute it for a specific scale in the inertial range, which we choose to be the Taylor scale $\lambda^M$.  Fig.~\ref{f_ratio_Re} shows how  $R_{\tau}(\lambda^M)$ changes with Reynolds number for each class of initial conditions.  We see that this ratio tends towards a constant whose value depends on the type of flow, although higher Reynolds numbers should be investigated to confirm this tendency.  The transition to a converged value of $R_\tau$ appears to take place at $Re \approx 3\times 10^3$, after a long evolution over a sequence of smaller Reynolds numbers.  For this reason, simulations at moderate Reynolds numbers can only tentatively show the asymptotic results needed to understand fully developed turbulence.
  
  \begin{figure}[htb]
\includegraphics[width=7cm, height=50mm]{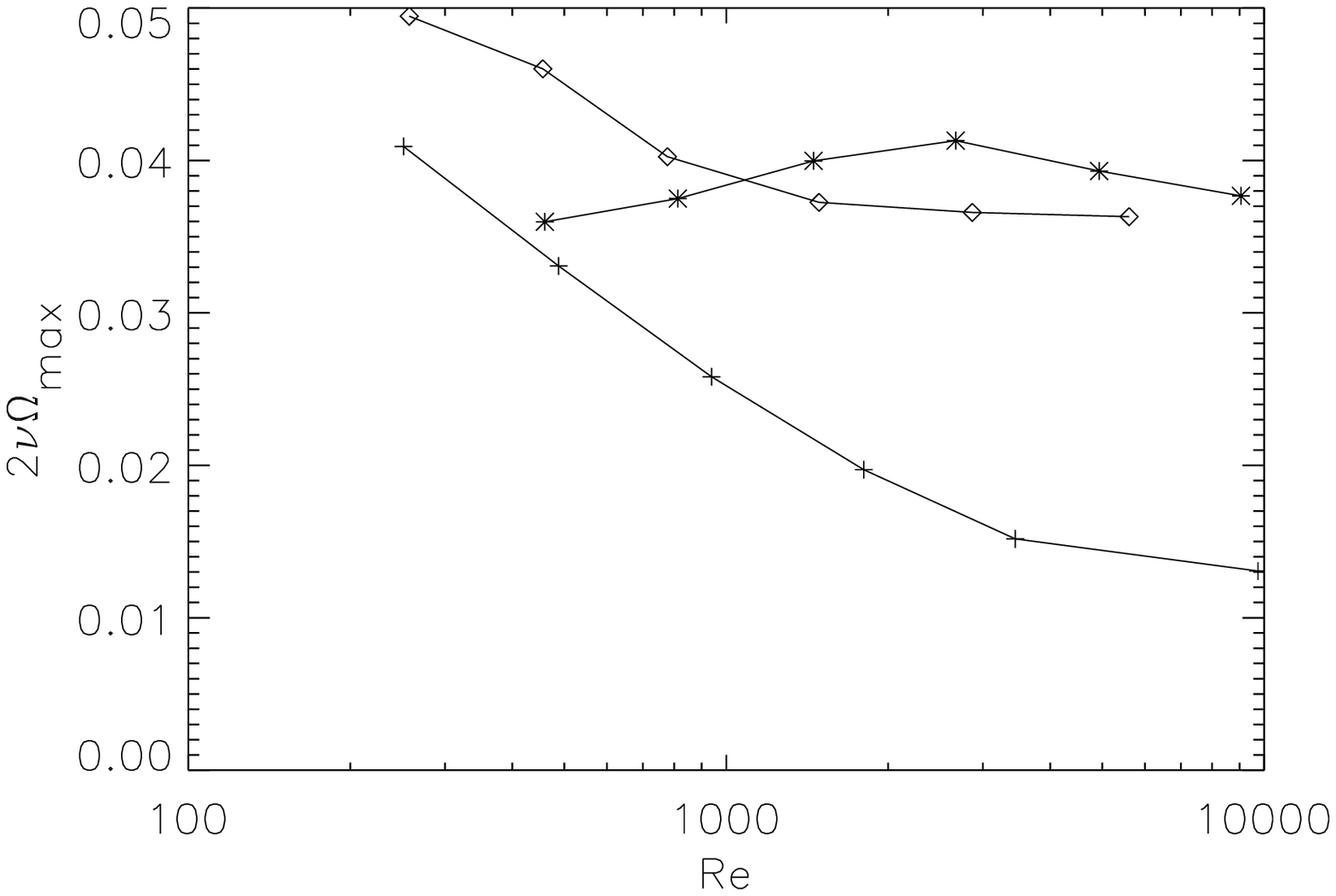}\\
\includegraphics[width=7cm, height=50mm]{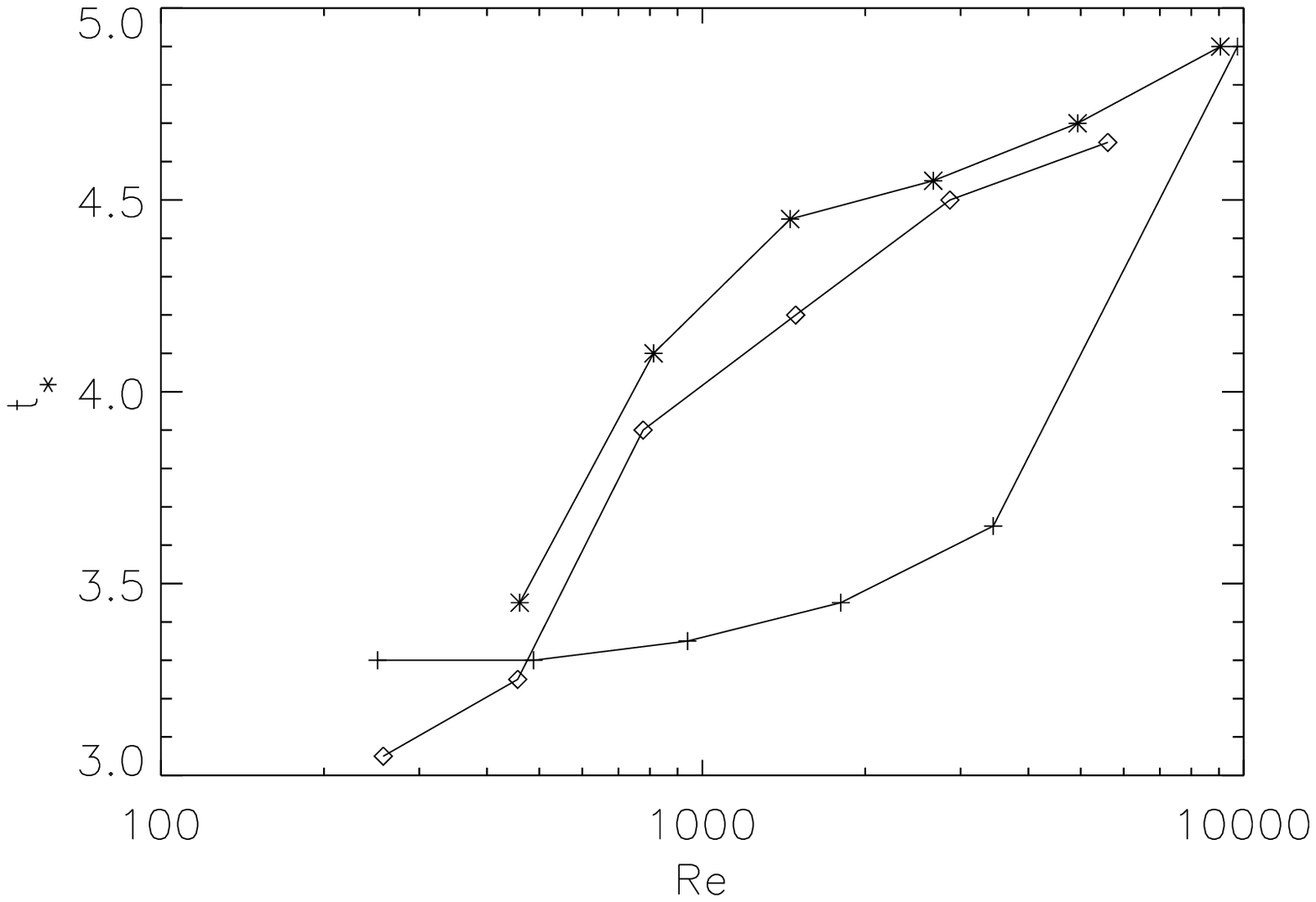}\\
\includegraphics[width=7cm, height=50mm]{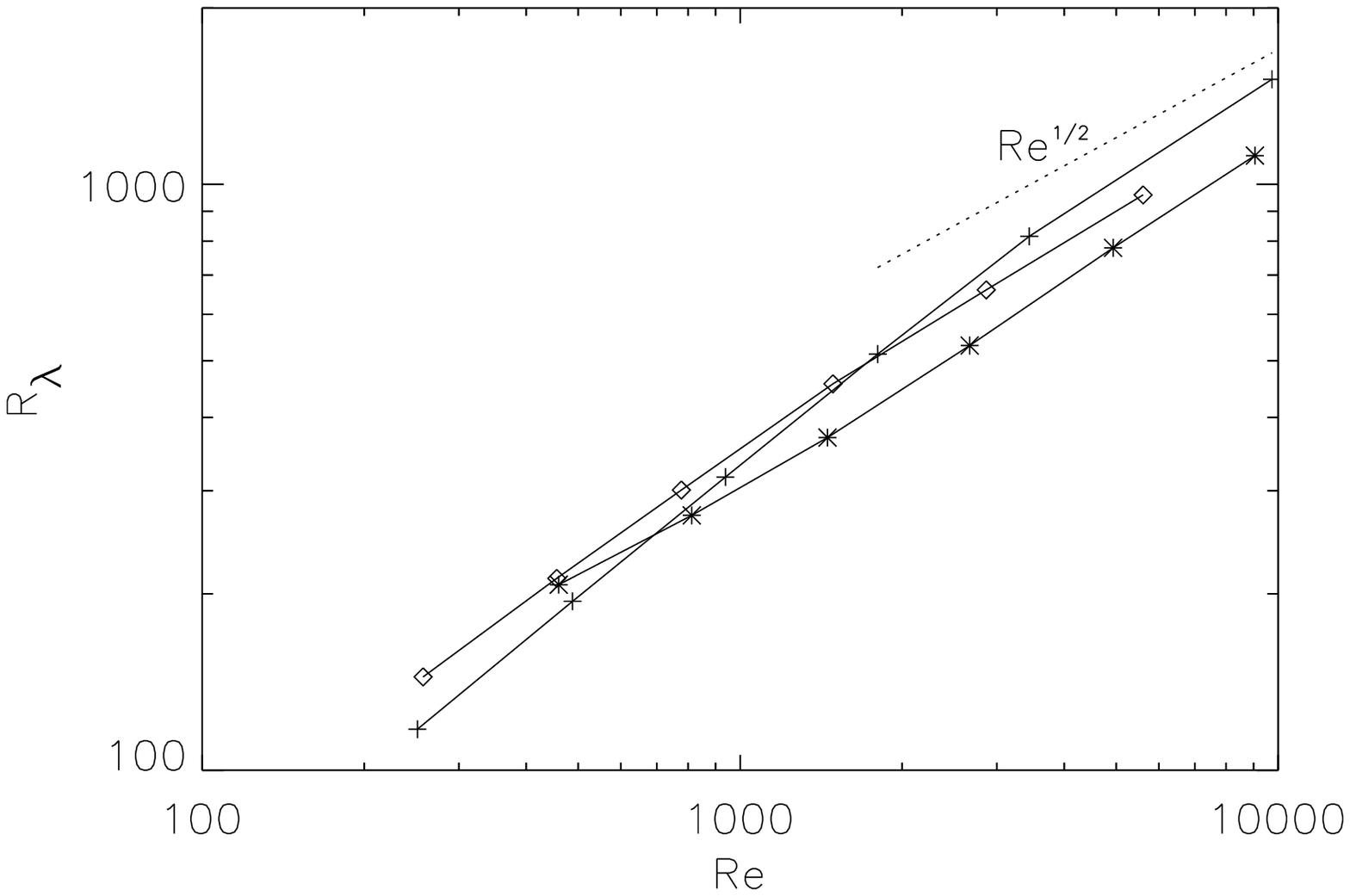}
\caption{ 
Scaling as a function of Reynolds number $Re$ of the maximum value of the dissipation over time, (\textit{top}), of the time at which this maximum is reached (\textit{middle}) and of the Taylor Reynolds number $R_{\lambda}$ computed at the instantaneous peak of dissipation (\textit{bottom}); the straight line indicates the turbulent scaling $R_\lambda \sim Re^{1/2}$. Symbols as in Fig.~\ref{f_ratio_Re}.}
\label{f_scaling_diss_taylor} \end{figure}

\begin{figure}[htb]
\includegraphics[width=7cm, height=50mm]{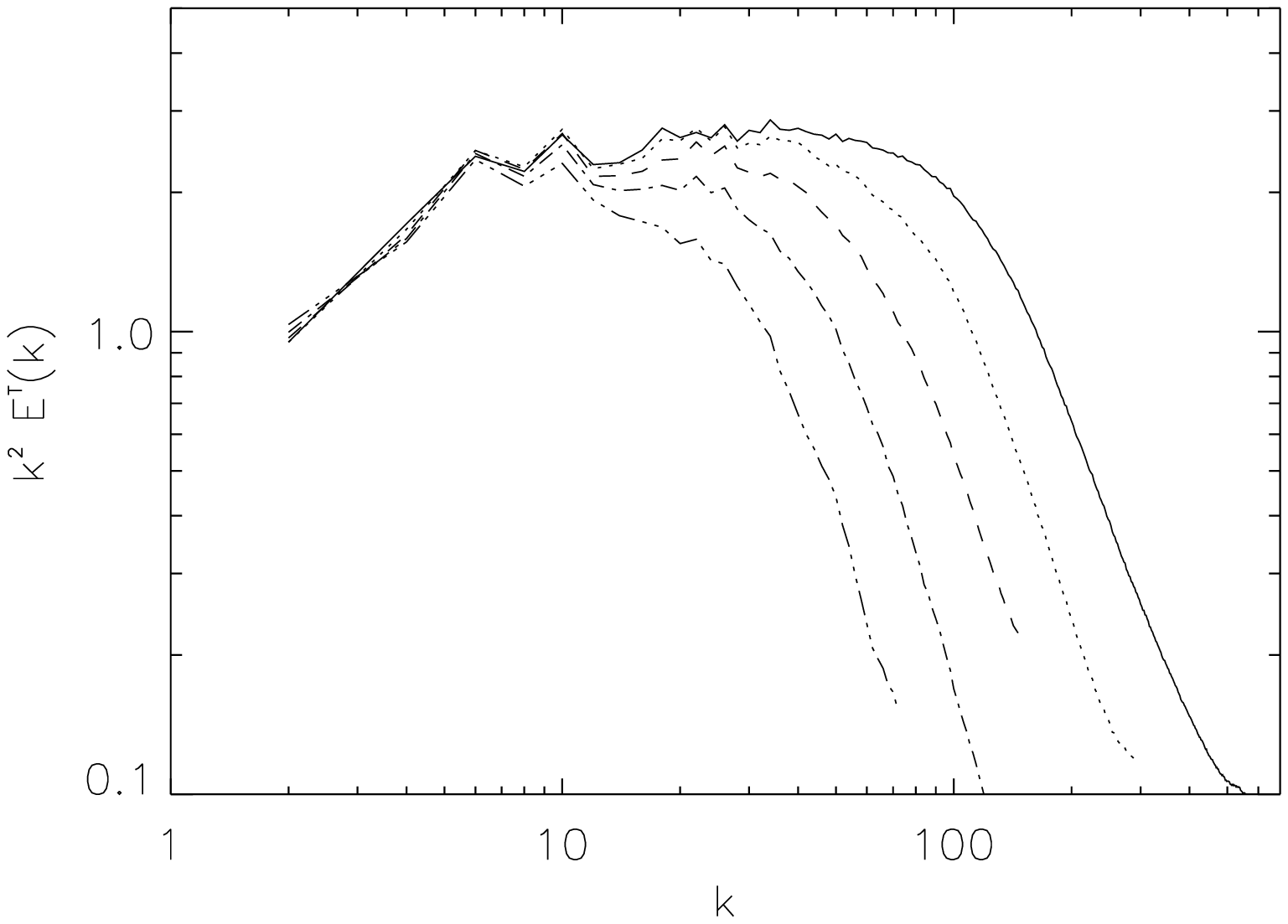}\\
\includegraphics[width=7cm, height=50mm]{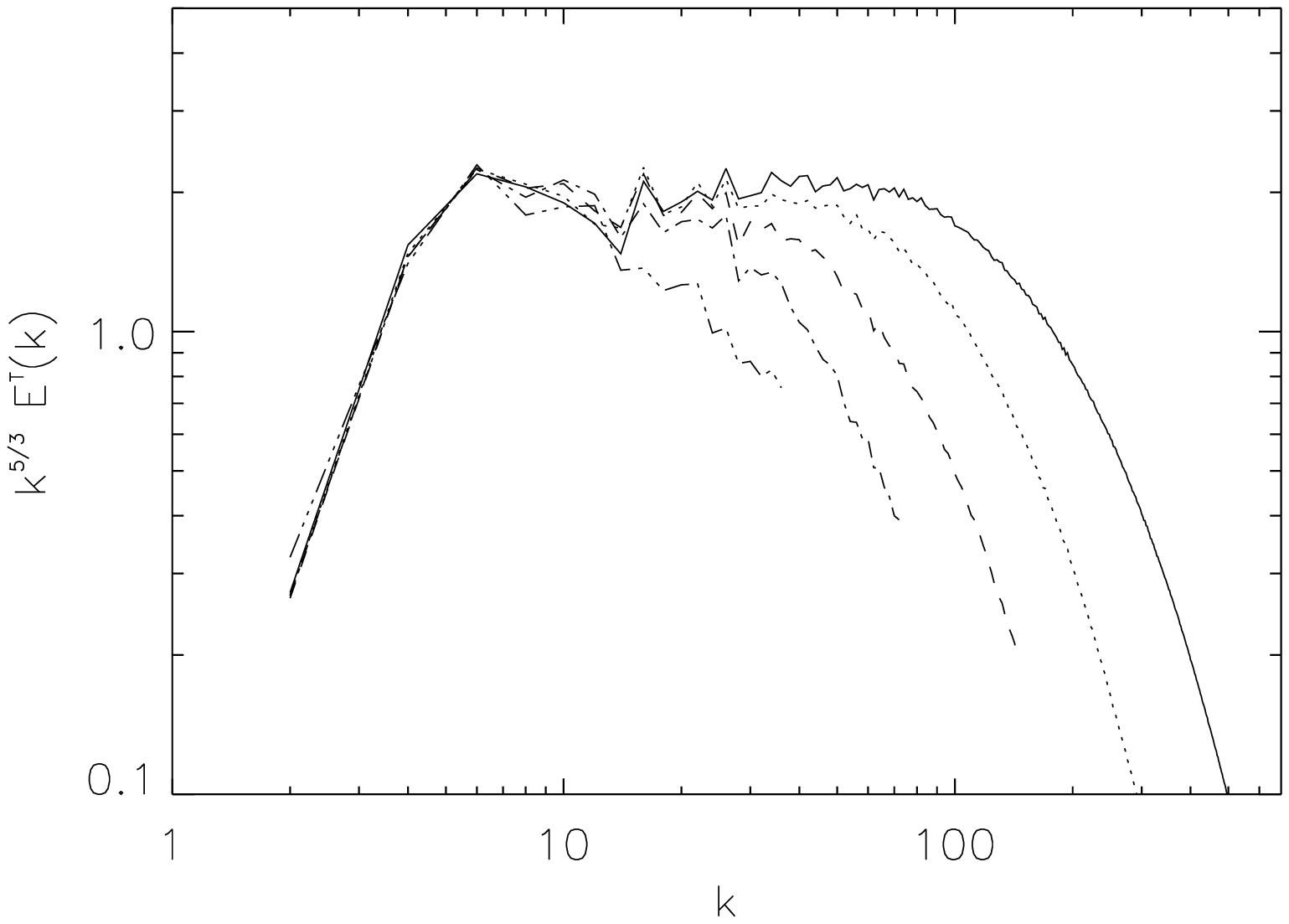}\\
\includegraphics[width=7cm, height=50mm]{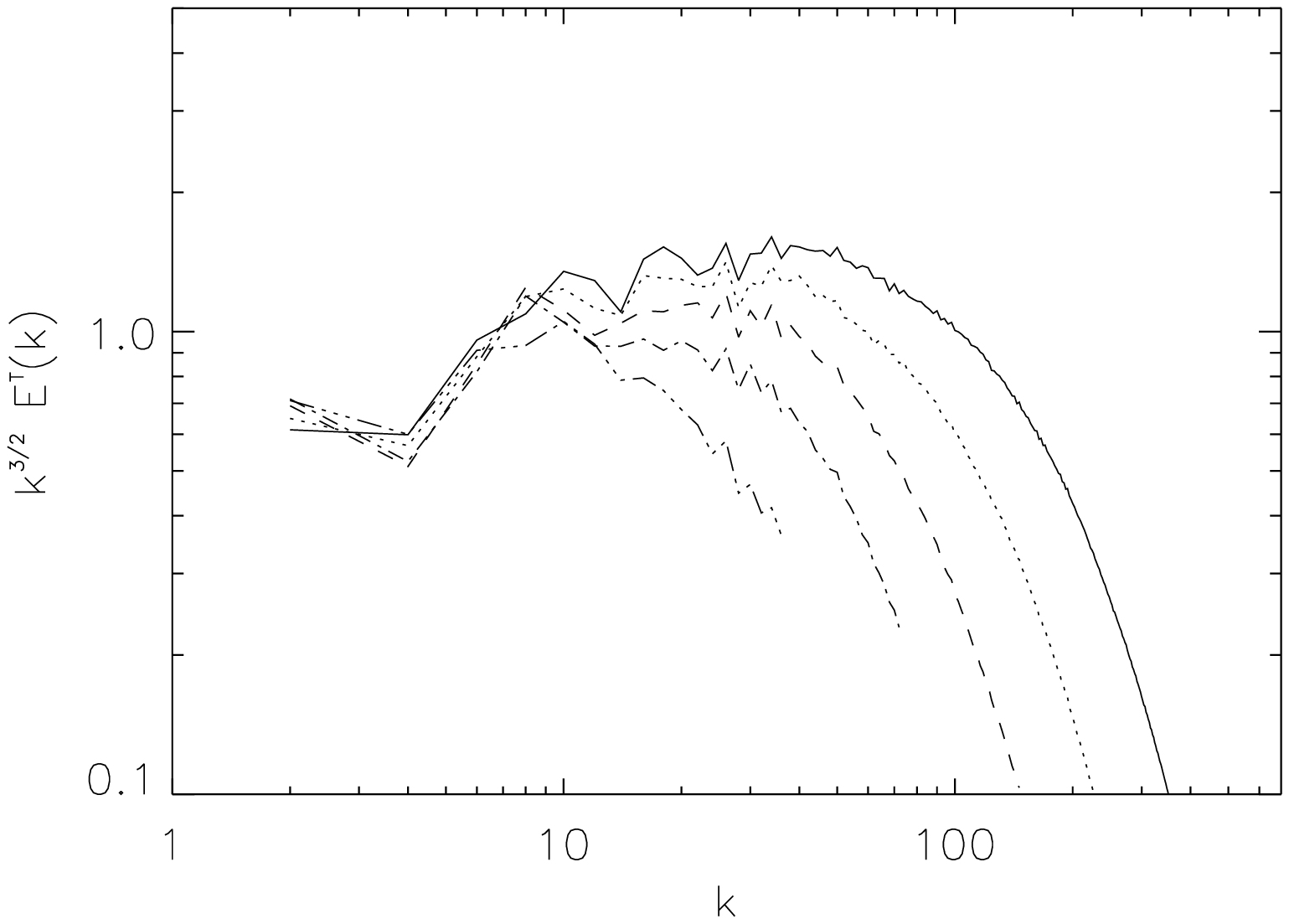}
\caption{
Total energy spectra averaged over $\Delta t = 0.5$ (1.5 to 2 turnover times) around the time of maximum dissipation for different Reynolds numbers for the following flows (see Table \ref{table1}): I runs compensated by $k^2$ (\textit{top}),  A runs compensated by $k^{5/3}$ (\textit{middle}) and C runs compensated by $k^{3/2}$ (\textit{bottom}). Dash-triple dots, dash-dots, dashes, dots, and solid lines represent respectively the runs I2 to I6, A2 to A6 and C2 to C6. Equivalent resolutions range from $128^3$ to $2048^3$ grid points.}
\label{f_comp_spec_3_av} \end{figure}

Other indicators can be examined to further suggest convergence, at least for the highest two resolutions (runs 5 and 6) for each class of flow.  In Fig.~\ref{f_scaling_diss_taylor} we show---again as a function of Reynolds number---the variation of the maximum of dissipation (\textit{top}), the time at which this maximum is reached (\textit{middle}), and the Taylor Reynolds number (\textit{bottom}).  We observe that the maximum of dissipation for the A and C initial conditions tends to level off towards a constant value as the Reynolds number is increased, as seen before in two-dimensional MHD \cite{biskamp89, politano89}, three-dimensional Navier-Stokes \cite{kaneda}, and an earlier three-dimensional MHD study \cite{1536a}.  The I flows seem to show a different trend, so higher resolution runs of this class (at higher Reynolds numbers) are necessary to determine if the maximum of dissipation does indeed reach a constant; here we can only observe a slowing of its decrease, and in fact the trend follows a power-law when examining the first maximum (as opposed to the absolute maximum, at $t\approx 3.25$ and $\approx 4.75$ respectively, see Fig. \ref{f_temp_enr_diss}, top and \cite{gafd}).  It should nevertheless be noted that, for each class, the maximum of dissipation occurs at a time that depends significantly on the Reynolds number (Fig.~\ref{f_scaling_diss_taylor}, \textit{middle}).  Finally, we observe a clear scaling of the Taylor Reynolds number $R_{\lambda}$ \cite{note2} with the large-scale Reynolds number $Re$ (\textit{bottom}):  the two flows at the highest Reynolds numbers for each class are consistent with $R_{\lambda}\sim Re^{1/2}$, indicating again  that we have reached an asymptotic state. 

With reasonable evidence of convergence of flows beyond a Reynolds number threshold, we can now turn to the energy spectra we observe for each class of flow.  Fig.~\ref{f_comp_spec_3_av} displays, for each class of initial conditions, the total energy spectra averaged again over an interval $\Delta t = 0.5$ about the maximum of dissipation of each run (see figure caption and Table \ref{table1}).  
The spectra are compensated by $k^{2}$ for the I flows, by $k^{5/3}$ for the A flows, and by $k^{3/2}$ for the C flows.  These plots strongly suggest that the scaling predictions of WT, K41, and IK give credible descriptions of the I (\textit{top}), A (\textit{middle}), and C (\textit{bottom}) flows, respectively, recognizing that intermittency can steepen the spectrum of the self-similar solutions.
We note also that for C6 the $k^{-3/2}$ scaling seems to end at the magnetic Taylor scale, beyond which bending of magnetic field lines is felt and a steeper power law is possible, as already observed for a general non-symmetric flow \cite{1536b}, but no such double power law is observed for the other two classes of flow.  Furthermore, a bottleneck at the beginning of the dissipation range is noticebly absent or undetectable, likely due to the intrinsic nonlocality of nonlinear interactions in MHD \cite{alex}.

Table \ref{table2} gives quantitative data of the characteristic scales in the highest resolution runs of each class, including the integral and Taylor scales, as well as the Taylor Reynolds number, based on $\lambda^T$.  When based on the magnetic Taylor scale $\lambda^M$ instead, the Taylor Reynolds numbers are 1600 (for I6), 950 (for A6), and 910 (for C6) respectively.



\begin{figure}[htb]
\includegraphics[height=50mm]{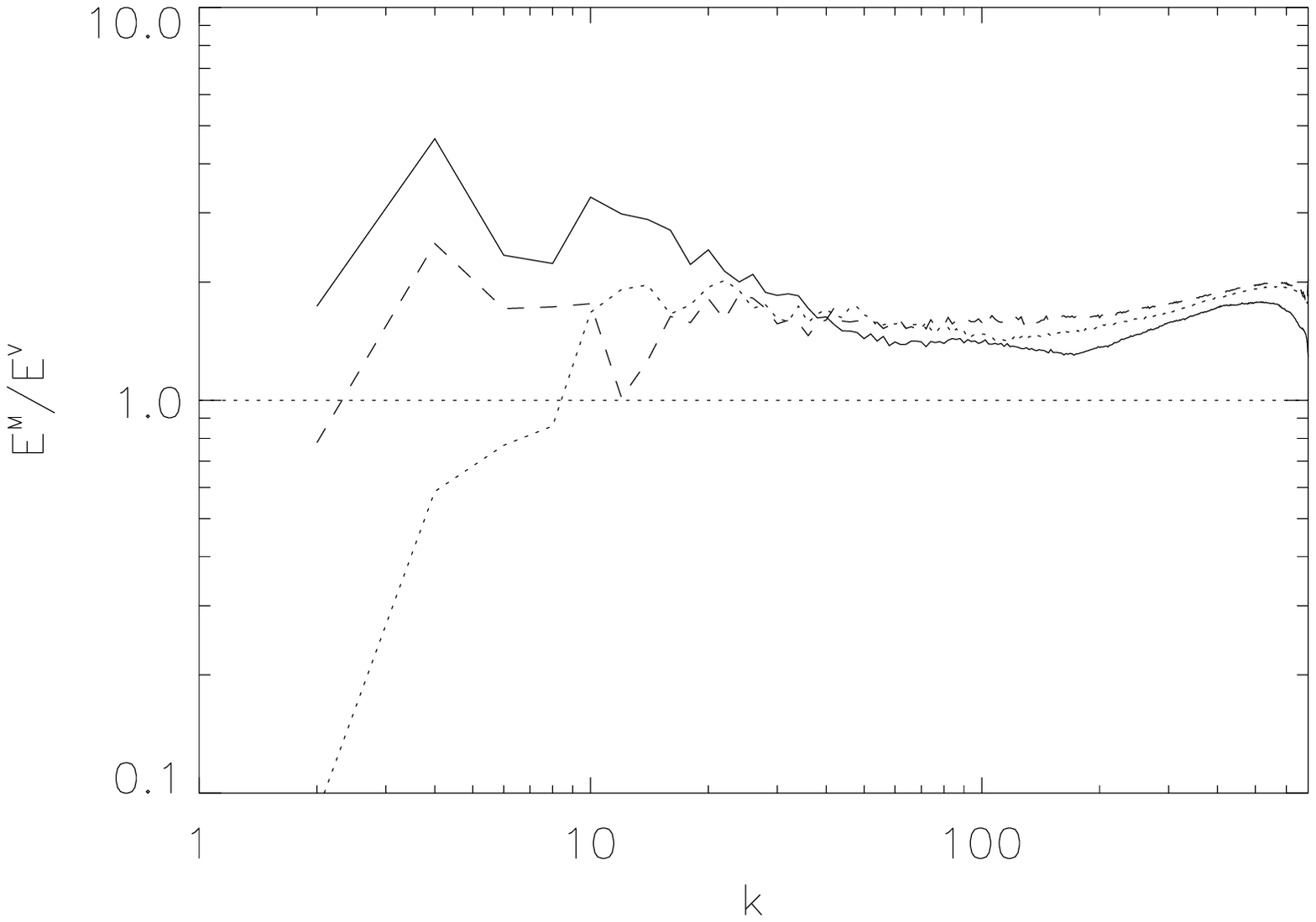}
\includegraphics[height=50mm]{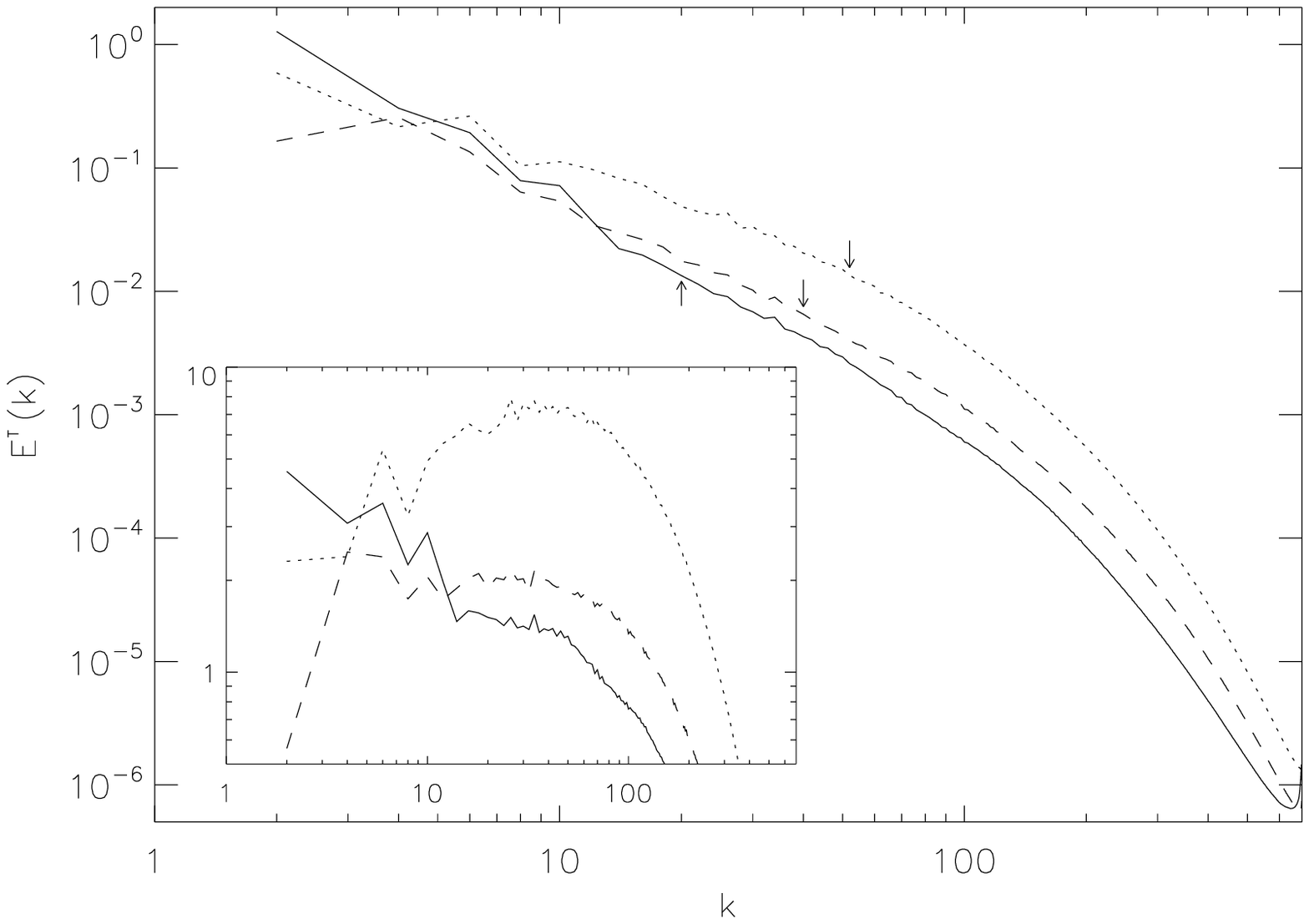}
\caption{
({\it Top}) Magnetic to kinetic energy ratio as a function of wavenumber at later time ($t \in [6,6.5]$) than in  Fig.~\ref{f_spec_ratio_enr} and averaged over the same length of time.  Again, we plot I6 (solid), A6 (dashed) and C6 (dotted).  Note that the excess of magnetic energy at low wavenumber in I6 has subsided to 1/15 of its former value.  ({\it Bottom}) Total energy spectra for the same runs, over the same time interval.  The arrows indicate the new magnetic Taylor wavenumber for each run.  In the insert, the same spectra are compensated by $k^{3/2}$.  Note the similar energy ratios and inertial range scaling for the three flows.
} 
\label{late_times} \end{figure}



\section{Discussion and Conclusion} \label{s:con}

There is a wealth of theoretical, phenomenological, observational, and numerical studies of energy spectra for MHD turbulence. Solar wind data has seemed for a long while to favor a K41 classical spectrum, but a puzzling recent result is that, in some cases, both IK- and K41-type power laws have been observed for the velocity and magnetic field \cite{podesta}; moreover, recent data on MHD turbulence in the plasma sheet using the CLUSTER suite of satellites seem to indicate that the inertial index of this flow varies but with more likelihood for a $-2$ law and to a lesser extent a $-1.6$ power law for the energy spectrum \cite{cluster}.   
The tendency toward K41 or IK dynamics has also been observed recently in several numerical simulations \cite{matthaeus_RMHD,mueller,mason_08} with different forcing functions (see also \cite{yoshida07}).  In fact, it has been shown numerically for the reduced MHD equations \cite{strauss} that a whole palette of spectra is possible \cite{matthaeus_RMHD,rapazzo_07}.

 While important theoretically for establishing how dynamical energetic exchanges in MHD turbulence occur, distinguishing amongst possible spectral indices numerically is difficult at best.  Equally unclear is whether one power-law solution exists in all cases or several classes of universality are possible \cite{pablo_NJP}, or if the resulting spectra can exhibit an arbitrary power law in a range constrained by integrability both at small and large scales, avoiding the so-called ultraviolet and infrared catastrophic divergences \cite{frisch_book}.  It was found recently that for a given set of initial conditions, an isotropic IK spectrum results in the large scales, with a transition at the magnetic Taylor scale towards a WT anisotropic spectrum, as mentioned earlier, with both ranges only moderately resolved \cite{1536b}.  Furthermore, boundary conditions and other forcing functions (e.g., random) may play a role as well.  For example, it was found in a pioneering paper that the spectrum (in reduced MHD) can become very steep when the time scale of the external forcing is varied \cite{matthaeus_RMHD}.

To this debate we contribute here our numerical confirmation that different energy spectra may indeed emerge in the specific case of decaying MHD turbulence in the absence of a uniform magnetic field.
For this purpose, we have generalized the Taylor-Green flow to MHD and studied three different configurations that lead to three different types of behavior, even though, from a statistical point of view, these configurations were a priori equivalent since they had the same invariants ($E^T$, $H^C$ and $H^M$) and the same equipartition between kinetic and magnetic energy at initial time.  By taking advantage of the natural symmetries of these flows, we have been able to examine higher Reynolds numbers than for a full DNS for a given cost; we could also perform a convergence study in terms of Reynolds number, with $R_\lambda \sim 960$ or above in all three flows at the highest resolution.  We found that the I flow behaves somewhat differently, with a slower dissipation of energy at a given Reynolds number and a lower maximum of dissipation, and that all three flows had different partition between the kinetic and magnetic energy at large scales, a partition that is likely at the origin of their different spectral behavior.

That the observed discrepancy in spectral behavior is not linked intrinsically to a given type of initial condition but rather correlated with the ratio for time scales as displayed in Fig.~\ref{f_spectra_enr_times} is confirmed by the following analysis.  It is evident in Fig.~\ref{f_temp_enr_diss} (\textit{bottom}) that at late time ($t \gtrapprox 5.5$) the ratio of magnetic to kinetic energy does not differ as much for the three types of flows studied in this paper.  Therefore, we examine in Fig.~\ref{late_times} the ratio of magnetic to kinetic energy spectra (\textit{top}) and the energy spectra (\textit{bottom}) averaged over $\Delta t = 0.5$, as before, but now at a later time beginning at $t=6$.  Indeed, when the kinetic and magnetic energy are comparable, 
the energy spectra of the I and A flows likewise have comparable spectral index, the C flow (dotted line) being somewhat shallower although at this late time and lower Reynolds number, the difference is hard to tell. However, the I flow is clearly not as steep as at earlier times.

The different power laws observed in this study can in principle be associated with known ``solutions'' (K41, IK, or WT) of MHD turbulence (omitting here intermittency corrections), and they are found to be correlated with the ratio of the nonlinear to the Alfv\'en time.  This is linked to the competition between nonlinear steepening and dispersion due to waves, which can interact as they propagate forwards and backwards along a mean field---here a local mean field, since there is no imposed background field.  Whether the attractors for MHD (associated with different spectral indices) are one, many, or infinite, remains to be determined.  It is conceivable that multiple fixed points can co-exist, linked with K41 (fluid-like), IK (balance between steepening and dispersion) and WT (turbulence and waves) dynamics.
For example, one numerical study, though performed at low resolution, showed that there are indeed several possible attractive solutions for decaying incompressible MHD in the absence of an imposed magnetic field:  one dominated by the velocity, another dominated by the magnetic field, and the third with quasi-equipartition between the two modes of energy \cite{matthaeus_3states}.  Are the solutions we observe in our study, which build up on a fast time scale of the order of the eddy-turnover time, associated with the slow dynamics of the attractive solutions of decaying MHD?  This hypothesis might be tested by performing multiple selective decay simulations for very long times for the three TG MHD flows analyzed in this paper and variants thereof, a task left for future work.

The initial conditions we have chosen are concentrated in the largest scales in order to obtain the largest Reynolds numbers possible for these flows.  Moving the characteristic wavenumber of the initial conditions to smaller scales would allow for an inverse transfer of magnetic helicity to occur, and this in turn may well influence the overall dynamics.  A substantial amount of cross helicity, a different initial ratio of magnetic to kinetic energy, or a non-unit magnetic Prandtl number could also have an important ciimpact on the resulting energy spectra.  For example, it has been known for a long time that global and local correlation of velocity and magnetic field fluctuations can directly influence the spectral indices of the Els\"asser variables and hence of all other spectra, with a steeper slope for the dominant mode.  These effects have been studied in the context of isotropic turbulence closures \cite{grappinm+}; in weak turbulence \cite{galtier00}; in two-dimensional simulations \cite{politano89,pof31}; and in analytical studies of small-scale structures and wave interactions \cite{XY,boldyrev}. Similarly, it would be of interest to study the forced case, using the three (I, A, C) configurations studied in this paper.


That the inertial index of energy spectra can vary over a discrete or continuous range of values therefore has been supported by many previous studies.  This lack of universality has been studied analytically in terms of eddy shape and alignment with magnetic fluctuations \cite{boldyrev}, numerically in terms of time scales in reduced MHD with forcing \cite{matthaeus_RMHD}, and experimentally in the different context of surface gravity wave turbulence \cite{lukaschuk}.  It has also been linked to the amplitude of forcing in another experimental study \cite{fauve}.
Although the regime we study here numerically is one at high Reynolds number, where the role of waves is blurred by their interactions with eddies and also by their nonlinear steepening within structures, we nevertheless consider it an open problem to know whether other spectra than the three presented here can be found in MHD turbulence.  Other open problems are to understand the interplay of eddies and waves in turbulence, particularly in the MHD case, and to unravel the role played by anisotropy induced by a large-scale magnetic field.

Whatever power law may exist for an energy spectrum, it may moreover be affected additionally by the amount of intermittency present in the flow.  Measurement of intermittency (as performed recently for active regions of the sun \cite{abramenko}) in our flows---and comparison among the three classes---is left for future work.  It should be linked to a study of the structures that develop, which both influence and are influenced by the nonlinear dynamics.  It is already known that current and vorticity sheets fold and roll-up at high Reynolds numbers \cite{1536b,gafd}. One further question is whether the set of anomalous exponents for the velocity and magnetic fields, or equivalently in terms of Els\"asser variables, which differ at second order corresponding to the energy spectrum, have a common asymptote at high order, probably determined by the structures with steepest gradients.


Further work should also include exploring runs at higher Reynolds numbers. Short of waiting for the next generation of resources (which will be made available to be developed, e.g. through the petascale computing initiative, one can resort to modeling methods, in addition to direct numerical simulation.  Akin to the one presented here, insofar as implementing a reduction of modes at a given Reynolds number, is the numerical algorithm that decimates modes (somewhat arbitrarily) in the dissipation range \cite{sparse}.  Another possibility is the use of large-eddy simulations that compare well against high-resolution direct numerical simulations, such as in \cite{julien1}.  Of a different nature is the Lagrangian averaging approach, or alpha model, which can be viewed as a sort of direct numerical simulation methodology imposing a filter to the small scales by means of a closure consistent with preserving the Hamiltonian nature of the flow, although these averaged equations conserve the ideal invariants using a different norm than ${\cal L}_2$ \cite{holm}.  With alpha models (see e.g. \cite{lamhd1}), it has recently been shown that the result found in \cite{1536b} of two inertial ranges for MHD, of isotropic IK followed by a weak turbulence spectrum, can be recovered at substantially lower cost.  Using a combination of such modeling tools may allow for parametric investigations of MHD turbulence and thereby lead to a better understanding of such flows as they occur in geospace, the heliosphere, and the interstellar medium, and their influence for example on cosmic ray propagation or on the solar-terrestrial interactions.

\vspace{1cm}

\begin{acknowledgments}

Computer time was provided through NSF MRI -CNS-0421498, 0420873 and 0420985, NSF sponsorship of NCAR, the University of Colorado, and a grant from the IBM Shared University Research (SUR) program. Ed Lee was supported in part from NSF IGERT Fellowship in the Joint Program in Applied Mathematics and Earth and Environmental Science at Columbia. PDM acknowledges financial support from the Carrera del Investigador Cient\'{\i}fico of CONICET.
\end{acknowledgments}



\begin{thebibliography}{99}
\bibitem{note1}
In two space dimensions, the slower decay of MHD turbulence, compared to Navier-Stokes flows, has been studied both with a model and with direct numerical simulations, see \cite{galtier_decay1}.
\bibitem{note2}
The Taylor Reynolds number, whose given value is computed at the peak of dissipation, saturates from that time onward.
\end{thebibliography}
\end{document}